\newif\ifsupp
\supptrue 

\documentclass[superscriptaddress,twocolumn,10pt,prl,floatfix,nofootinbib]{revtex4-2}

\input{glyphtounicode}
\usepackage[T1]{fontenc}
\usepackage[utf8]{inputenc}
\usepackage[english]{babel}
\usepackage{amsmath}  
\usepackage{amsfonts} 
\usepackage{graphicx} 
\usepackage[pdftex,bookmarks=true,bookmarksopen,bookmarksnumbered,
                colorlinks,
                linkcolor=blue,
                citecolor=blue]{hyperref}
\graphicspath{ {./} }
\usepackage{natbib}
\setcitestyle{super}
\usepackage{array}
\usepackage[table]{xcolor}
\usepackage{xcolor}
\usepackage{xr}
\usepackage{braket}
\usepackage{booktabs}
\usepackage{multirow}
\usepackage{caption}
\usepackage{dsfont} 

\usepackage{tabularx}
\usepackage{siunitx}

\usepackage{subcaption}
\usepackage{bigstrut}

\newcommand{\aref}[1]{\hyperref[#1]{Appendix~\ref*{#1}}}
\DeclareCaptionLabelSeparator{line}{\hspace{2pt}\bf{|}\hspace{2pt}}

\begin{document}

\captionsetup[table]{name={TABLE},labelsep=period,justification=centerlast,font=small}
\captionsetup[figure]{name={\bf{Figure}},labelsep=line,justification=centerlast,font=small}
\renewcommand{\equationautorefname}{Eq.}
\renewcommand{\figureautorefname}{Fig.}
\renewcommand*{\sectionautorefname}{Sec.}

\title{\color{black}Assessment of error variation in high-fidelity two-qubit gates in silicon\color{black}}


\author{Tuomo Tanttu}
\email{t.tanttu@unsw.edu.au}
\author{Wee Han Lim}
\affiliation{School of Electrical Engineering and Telecommunications, UNSW, Sydney, NSW 2052, Australia}
\affiliation{Diraq, Sydney, NSW, Australia}
\author{Jonathan Y. Huang}
\affiliation{School of Electrical Engineering and Telecommunications, UNSW, Sydney, NSW 2052, Australia}
\author{Nard Dumoulin Stuyck}
\author{Will Gilbert}
\affiliation{School of Electrical Engineering and Telecommunications, UNSW, Sydney, NSW 2052, Australia}
\affiliation{Diraq, Sydney, NSW, Australia}
\author{Rocky Y. Su}
\author{MengKe Feng}
\author{Jesus D. Cifuentes}
\author{Amanda E. Seedhouse}
\affiliation{School of Electrical Engineering and Telecommunications, UNSW, Sydney, NSW 2052, Australia}
\author{Stefan K. Seritan}
\author{Corey I. Ostrove}
\author{Kenneth M. Rudinger}
\affiliation{Quantum Performance Laboratory, Sandia National Laboratories, Albuquerque, NM and Livermore, CA, USA.}
\author{Ross C. C. Leon}
\altaffiliation[]{current address: Quantum Motion Technologies Ltd'}
\affiliation{School of Electrical Engineering and Telecommunications, UNSW, Sydney, NSW 2052, Australia}
\author{Wister Huang}
\altaffiliation[]{current address: ETH Zurich}
\affiliation{School of Electrical Engineering and Telecommunications, UNSW, Sydney, NSW 2052, Australia}
\author{Christopher C. Escott}
\affiliation{School of Electrical Engineering and Telecommunications, UNSW, Sydney, NSW 2052, Australia}
\affiliation{Diraq, Sydney, NSW, Australia}
\author{Kohei M. Itoh}
\affiliation{School of Fundamental Science and Technology, Keio University, Yokohama, Japan}
\author{Nikolay V. Abrosimov}
\affiliation{Leibniz-Institut für Kristallzüchtung, 12489 Berlin, Germany}
\author{Hans-Joachim Pohl}
\affiliation{VITCON Projectconsult GmbH, 07745 Jena, Germany}
\author{Michael L. W. Thewalt}
\affiliation{Department of Physics, Simon Fraser University, British Columbia V5A 1S6, Canada}
\author{Fay E. Hudson}
\affiliation{School of Electrical Engineering and Telecommunications, UNSW, Sydney, NSW 2052, Australia}
\affiliation{Diraq, Sydney, NSW, Australia}
\author{Robin Blume-Kohout}
\affiliation{Quantum Performance Laboratory, Sandia National Laboratories, Albuquerque, NM and Livermore, CA, USA.}
\author{Stephen D. Bartlett}
\affiliation{Centre for Engineered Quantum Systems, School of Physics, University of Sydney, Sydney, New South Wales 2006, Australia}
\author{Andrea Morello}
\affiliation{School of Electrical Engineering and Telecommunications, UNSW, Sydney, NSW 2052, Australia}
\author{Arne Laucht} 
\author{Chih Hwan Yang}
\author{Andre Saraiva}
\author{Andrew S. Dzurak}
\email[]{a.dzurak@unsw.edu.au}
\affiliation{School of Electrical Engineering and Telecommunications, UNSW, Sydney, NSW 2052, Australia}
\affiliation{Diraq, Sydney, NSW, Australia}
\date{\today}

\begin{abstract}
\color{black}
\textbf{Achieving high-fidelity entangling operations between qubits consistently is essential for the performance of multi-qubit systems and is a crucial factor in achieving fault-tolerant quantum processors. Solid-state platforms are particularly exposed to errors due to materials-induced variability between qubits, which leads to performance inconsistencies. Here we study the errors in a spin qubit processor, tying them to their physical origins. We leverage this knowledge to demonstrate consistent and repeatable operation with above 99\% fidelity of two-qubit gates in the technologically important silicon metal-oxide-semiconductor (SiMOS) quantum dot platform. We undertake a detailed study of these operations by analysing the physical errors and fidelities in multiple devices through numerous trials and extended periods to ensure that we capture the variation and the most common error types. Physical error sources include the slow nuclear and electrical noise on single qubits and contextual noise. The identification of the noise sources can be used to maintain performance within tolerance as well as inform future device fabrication. Furthermore, we investigate the impact of qubit design, feedback systems, and robust gates on implementing scalable, high-fidelity control strategies. These results are achieved by using three different characterization methods, we measure entangling gate fidelities ranging from 96.8\% to 99.8\%. Our analysis tools identify the causes of qubit degradation and offer ways understand their physical mechanisms.  These results highlight both the capabilities and challenges for the scaling up of silicon spin-based qubits into full-scale quantum processors.}
\color{black}

\end{abstract}

\maketitle

\begin{figure*}[ht!]
    \includegraphics[width = 1\textwidth,width=\textwidth, trim = 0cm 17cm 0cm 1cm]{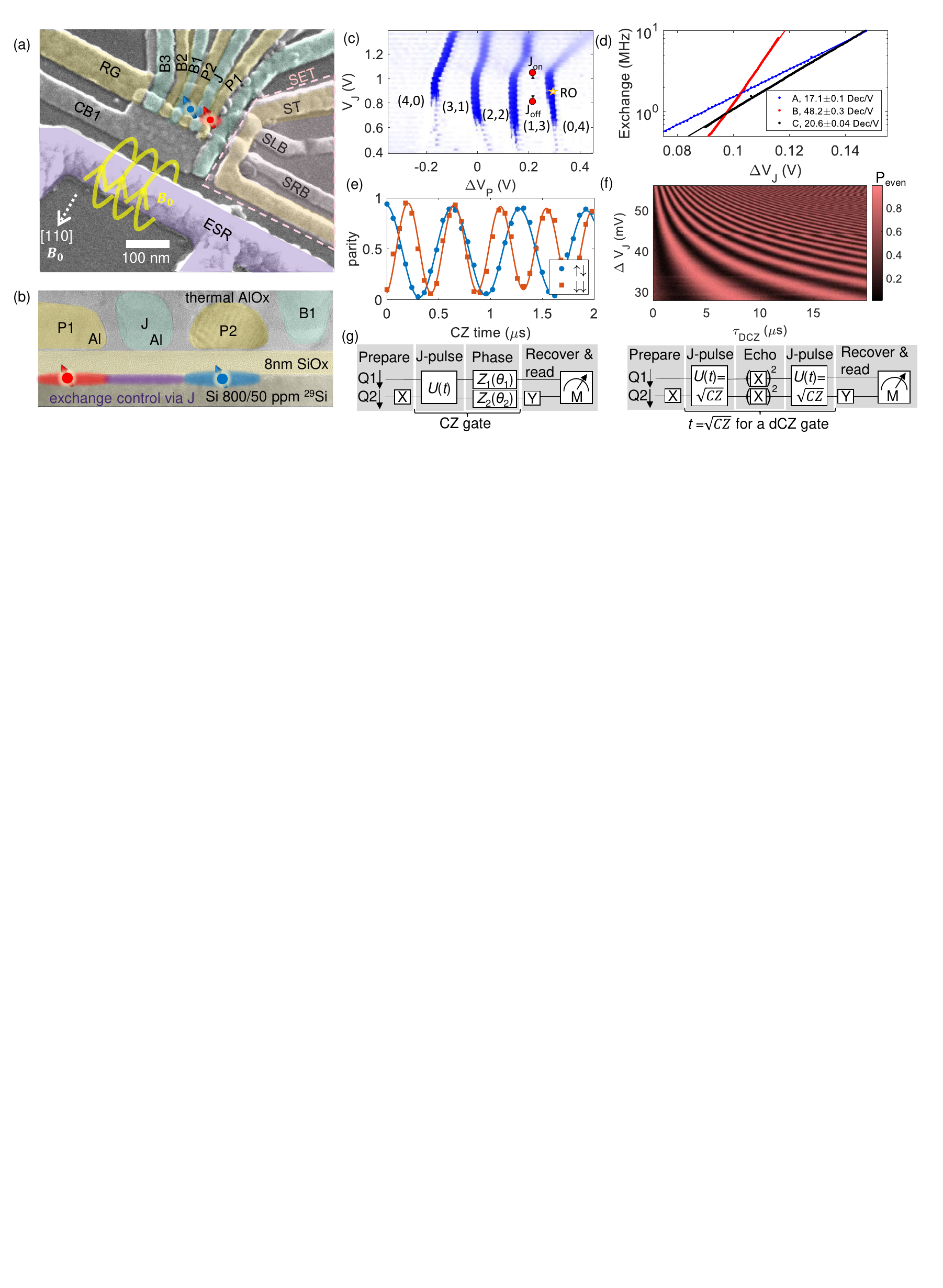}
    \caption{\textbf{Electrostatic quantum dots with tunable exchange.} 
    \textbf{a}, False color scanning electron micrograph (SEM) of a device similar to A and B.
    \textbf{b}, False color transmission electron micrograph (TEM) of a cross-section of a device similar to A and B. Shaded area shows the extension of electron wave functions under each dot. 
    \textbf{c}, Charge stability map of device A in isolated mode together with the important operation points for qubit operation ($J_{\textrm{on}}$ and $J_{\textrm{off}}$) and readout (RO). 
    \textbf{d}, Exchange energy for devices A, B and C as a function of $V_J$. The rate of increase of exchange is shown in the legend.
    \textbf{e}, CZ oscillations using the measurement sequence in \textbf{g} with two different control qubit initializations at a fixed level of~J.
    \textbf{f}, Oscillations of DCZ gate sequence as a function of J-gate pulse time and voltage level.
    \textbf{g}, Pulse sequences used in the experiments for CZ (left) and DCZ (right) in \textbf{e} and \textbf{f}.~\color{black}Here X-gate refer to $\textrm{X}^\frac{\pi}{2}$ to the corresponding qubit.\color{black}
 }
    \label{fig:sem}
\end{figure*}
The fidelities of qubit operations need to be consistent over time and across different qubits in order to achieve complex quantum computations, such as those in recent groundbreaking research~\cite{Google2019supremacy, Google2023error, USTC2020supremacy, Postler2022faulttolerantion, Xanadu2022supremacy}. The physical mechanisms behind the entanglement between qubits are key to the success of a two-qubit gate, and have a large impact in the performance of algorithms and error correction schemes~\cite{fowler2012surface, Stace2009threshold, Stace2010errorcorrection, Auger2017Fault}. Exchange-based entangling gates between silicon spin qubits have only recently matured enough to achieve high-fidelity operation~\cite{madzik2022precision, noiri2022fast, xue2022quantum, mills2022twoqubit, Weinstein2023universal}. 

\color{black}The variation of the operational parameters is particularly important in the case of spin qubits due to their nanometric physical size and nanosecond scale operation time.\color{black}~While all solid-state systems are subjected to materials noise and disorder, spin qubits probe these imperfections nearly at the atomic scale. In the case of SiMOS quantum dot qubits~\cite{Veldhorst2014addressable} this is especially pronounced since the qubits are pressed against the amorphous Si/SiO$_2$ interface. The issue of the consistency of high-fidelity operations then becomes central to translating this fabrication scalability into a qubit control scalability.

Our goal is to analyze the statistical characteristics and temporal stability of primitive two-qubit gate operations involving electron spins in SiMOS quantum dots, which is based on the Heisenberg exchange interaction. Our entangling gates are based on controlling the exchange between spins of electrons in neighbouring quantum dots by pulsing the height of the tunnel barrier between dots with an interstitial exchange-control electrode. We perform entangling gates with two strategies: a simple square pulse of voltage, which leads to an effective Ising interaction between spins implementing controlled phase (CZ)~\cite{Veldhorst20152Q}; or a composite gate consisting of two voltage pulses separated by a microwave pulse that performs single-qubit dynamical decoupling, referred to as decoupled controlled phase (DCZ) gates~\cite{Watson2018Nature}. We analyse the errors introduced by each of these strategies in detail, leveraging three state-of-the-art methods of validation.

Besides the multiple validation methods, we verify the consistency of the two-qubit gate operations by reproducing them in three different devices. Two devices (A and B) are nominally identical three-dot chains, having been fabricated in the same batch, with the gate layout shown in Fig.~\ref{fig:sem}a. The third device (C in Extended Fig.~\ref{extFig:StabilityMaps}b) has four dots instead, but the same choice of material stack (with aluminium gates and thermally grown Al${_2}$O${_3}$). All experiments shown here are based on forming only two of the dots at a time.

The level of isotopic purification of the silicon substrate is also different -- 800~ppm $^{29}$Si for devices A and B and 50~ppm for device C. We do not present results comparing silicon substrate purification levels, but qualitatively the more purified device C was operated with high fidelity without as much need for active feedback on the qubit parameters (only a single parameters, instead of 9 or 7 in the cases of devices A and B, respectively). 

Our electron spin qubits reside under the plunger gates, P1 and P2, and the oscillating $B_1$ field from a nearby antenna drives the electron spin resonance (ESR) induced by magnetic field, $B_0$. The exchange interaction is tuned with the voltage on the interstitial exchange gate, J. For measurement, we use a single electron transistor (SET) which senses the charge movement, which becomes conditional on the spins when we try to move both electrons into the same dot due to the Pauli exclusion principle. The detailed cross-section image in Fig.~\ref{fig:sem}b~\color{black}displays the active area, where the qubits are formed, revealing also the intrinsic oxide variation and fabrication inconsistencies that lead to variation in the device parameters and performance that lead to the difficulty in obtaining consistent fidelities across devices~\cite{cifuentes2023bounds}.~\color{black}The differences between devices are laid out in Extended Table~\ref{extTable:DeviceComparison}.

\begin{figure*}[ht!]
    \includegraphics[width = 1\textwidth,width=\textwidth, trim = 0cm 12.3cm 0cm 1cm]{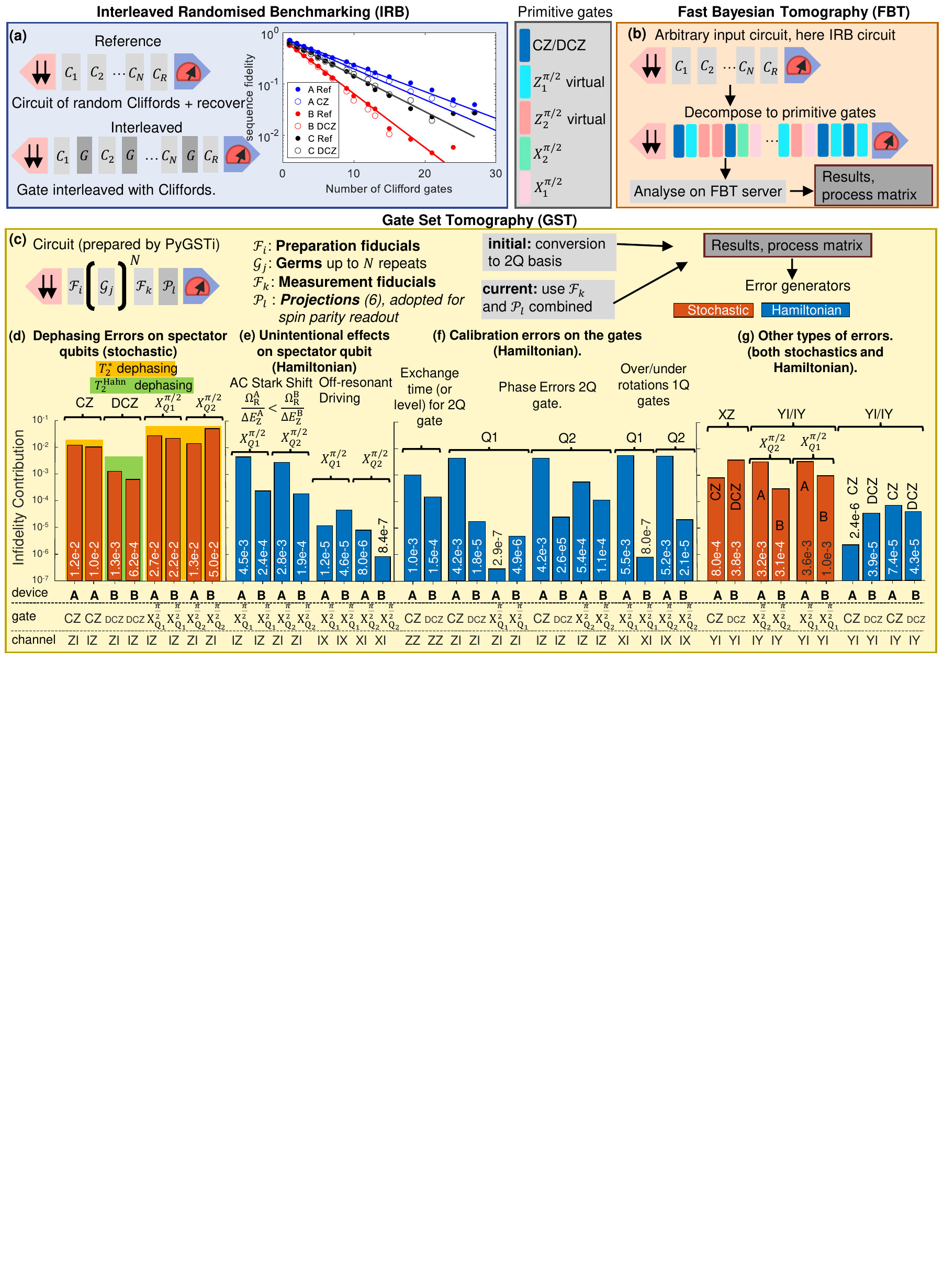}
    \caption{\textbf{Summary of Tomographic methods and selection of major identified physical error sources of the gate implementations based on GST for one and two qubit gates in devices A and B.} 
    \textbf{a}, Measurement sequence principle used in randomised benchmarking. The gate of interest is interleaved with random Clifford gates that are composed of five primitive gates each. Recovery probability of the randomised benchmarking sequence as a function of Clifford gates both for interleaved and reference sequences in all devices.
    \textbf{b}, Simplified fast Bayesian tomography workflow from experiment to result. FBT can analyse any gate sequence, IRB is used here as an example.
    \textbf{c}, Gate set tomography workflow used in our experiment.~\color{black}
    \textbf{d}, Major error generators of single qubit gates, Dephasing errors. These stochastic errors include the noise due to the $T_2^*$ like decay during the operation and also non-Markovian contextual error sources (see Fig.~\ref{fig:contextual}). The reduced error rates between DCZ is due to noise limited by $T_2^\textrm{Hahn}$ like decay instead.
    \textbf{e}, Physical errors which are Hamiltonian and occurring because of the operation such as AC-Stark shift, off-resonant driving or residual exchange during the single qubit operation.
    \textbf{f}, Calibration (systematic) errors, due to the errors in the calibration of the gates.
    \textbf{g}, Other major errors with no major physical attribution.\color{black}
 }
    \label{fig:MajorErrors}
\end{figure*}

The isolated mode~\cite{Eenink2019isolated,yang2020operation} stability map in Fig.~\ref{fig:sem}c for device~A with four electrons reveals the charge transitions between dots and allows us to choose the operational points in the device (see Extended Figs.~\ref{extFig:StabilityMaps}a and~c for~B and~C).~\color{black} We choose a symmetric operation point for our exchange on voltage to ensure that our two-qubit gate is not sensitive to the detuning noise. The dc-biases are typically chosen so that exchange on point is at high end of our dynamic range (400~mV) so that we can lower down the tunnel rates for the blockade readout. 

The spins in all our experiments are read out through the relative parityof the two spins using Pauli spin blockade for spin-to-charge conversion~\cite{Johnson2005blockade, seedhouse2021pauli}. After finding the resonance frequencies of the qubits, we analyse their coherence, Rabi frequencies and noise spectra. In one case, for device A, we use a vector magnet to study the effect of the direction of the constant magnetic field and identify the microscopic origin of the decoherence. Our analysis shows that the $T_2^*$ A is limited by both spin-orbit and $^{29}$Si noise~\cite{cifuentes2023impact}. Device B is likely to have similar noise limitations since it comes from the same fabrication batch, whereas device C is likely limited mostly by electric noise, thanks to its superior isotopic purity. Our $T_2^{\textrm{Hahn}}$ is more strongly limited by the charge rather than hyperfine noise in all devices. To fight the low frequency components of the intrinsic $1/f$ noise and slow jumps due to hyperfine interactions with $^{29}$Si nuclear spins, we apply frequency feedback to devices A and B to keep the microwave control in resonance~\cite{stuyck2023realtime}. This is not necessary for device C.

To run high fidelity two-qubit gates without being limited by the residual exchange at the nominally off state, we aim at an exchange swing that provides preferably at least a $10^4$ ratio between on and off states, which is achievable for all devices with our 400~mV dynamical range in the voltage of the J~gates, as shown in Fig.~\ref{fig:sem}d.\color{black}

\subsection{Multi-qubit physical errors}
\color{black}
We utilise three different validation methods -- interleaved randomised benchmarking (IRB)\cite{Knill2008RBM,Magesan2012IRB}, fast Bayesian tomography (FBT)\cite{Evans2022FastBayesian}, and gate set tomography (GST)\cite{greenbaum2015introduction, Nielsen2021gatesettomography} -- illustrated in Figs.~\ref{fig:MajorErrors}a, \ref{fig:MajorErrors}b, and \ref{fig:MajorErrors}c, summarised in Extended Table~\ref{ExtTable:QCVVmethods}, and described in further detail in the Supplementary Discussion: Comparison with other high fidelity two spin systems in silicon. \color{black}Our tomographic analyses allow us to identify the underlying quantum process and form conjectures about the error physical mechanisms. We explored not only the entangling gates, but also the two-qubit processes generated by single qubit gates, which elucidates effects such as the exposure of a qubit to decoherence when idling while the other qubit is being controlled, and the effect of crosstalk as well as contextual errors.

We run a gate set tomography (GST) experiment in device~A with CZ implementation and in device~B with DCZ implementation.~\color{black} We generate the circuits for the analysis with special consideration for the measurement effects of the parity readout, as opposed to the more common readout of individual qubits. Two strategies are adopted to that end, one entailing the addition of projections of single qubit states into the parity and repeating the experiments for these projections, and the other by adapting the analysis tool itself (pyGSTi~\cite{Nielsen2021gatesettomography}) to handle the measurement effects directly at the analysis stage. Incorporating the native measurement operation explicitly and generating measurement fiducials accordingly is found to be the most efficient strategy, see more details in the supplementary discussion: Gate set tomography with parity readout.

The most immediate form of analysis of the errors detected by GST is the breakdown between Hamiltonian and stochastic parts, which can be done through a mathematical framework without assumptions of the underlying mechanisms~\cite{blume2022taxonomy}. The Hamiltonian part of the error is in general associated with calibration issues, unintended Hamiltonian terms or forms of contextual error that only become evident once long circuits are performed (such as the consistent heating of the chip due to the long string of microwave pulses). Feedback into the gate control parameters allows us to minimize these errors until we achieve consistently high fidelities.

Removing these peculiarities of the qubit performance from the picture, the infidelity becomes dominated by stochastic effects that we can compare across devices (note that the infidelity contribution of Hamiltonian errors is naturally weaker, with a quadratic dependence, compared to the linear dependence on stochastic errors). In Figs.~\ref{fig:MajorErrors}d, \ref{fig:MajorErrors}e, \ref{fig:MajorErrors}f, and \ref{fig:MajorErrors}g we have grouped the major error channels resulting from this optimization loop according to their physical interpretation; the dephasing errors, calibration errors, physical errors, and other uncategorised errors, respectively. 

One striking difference between the CZ and DCZ implementations is that both the stochastic and Hamiltonian IZ and ZI errors are significantly smaller for DCZ. By its design, DCZ echoes out the phase accumulation due to the quasistatic shifts in the spin Larmor frequencies, and hence suppresses stochastic IZ and ZI errors. It also suppresses the Hamiltonian~IZ and~ZI errors incurred due to the spurious Stark shift created by the voltage pulse on the exchange gate. This is the main reason our DCZ gates perform better in device~B even though the coherence times $T_2^*$ are worse for both qubits in this device compared to device~A. This is the primary contributor to differences in the fidelities between the two implementations.

The unintentional driving errors in Fig.~\ref{fig:MajorErrors}e are created by frequency crosstalk. The difference in qubit frequencies $\Delta E_Z$ is large compared to the Rabi frequencies $\Omega_R$, but is still relevant when high fidelity operation is attempted, generating AC Stark shift (Z~error) and off-resonant driving (X~error). These errors appear as Hamiltonian since they are systematic. Calibration errors in Fig.~\ref{fig:MajorErrors}f are also Hamiltonian and are caused by the realistic limits in calibration accuracy of the gates, leading to errors such as under/over rotation of the gates, exchange level errors.
\color{black} We also note that the environment circumstances in which calibration is performed are different from those during the circuits, which leads to contextual errors.

\begin{figure}
    \includegraphics[width = 1.5\columnwidth, angle = 0,trim = 0cm 17.3cm 4cm 0cm]{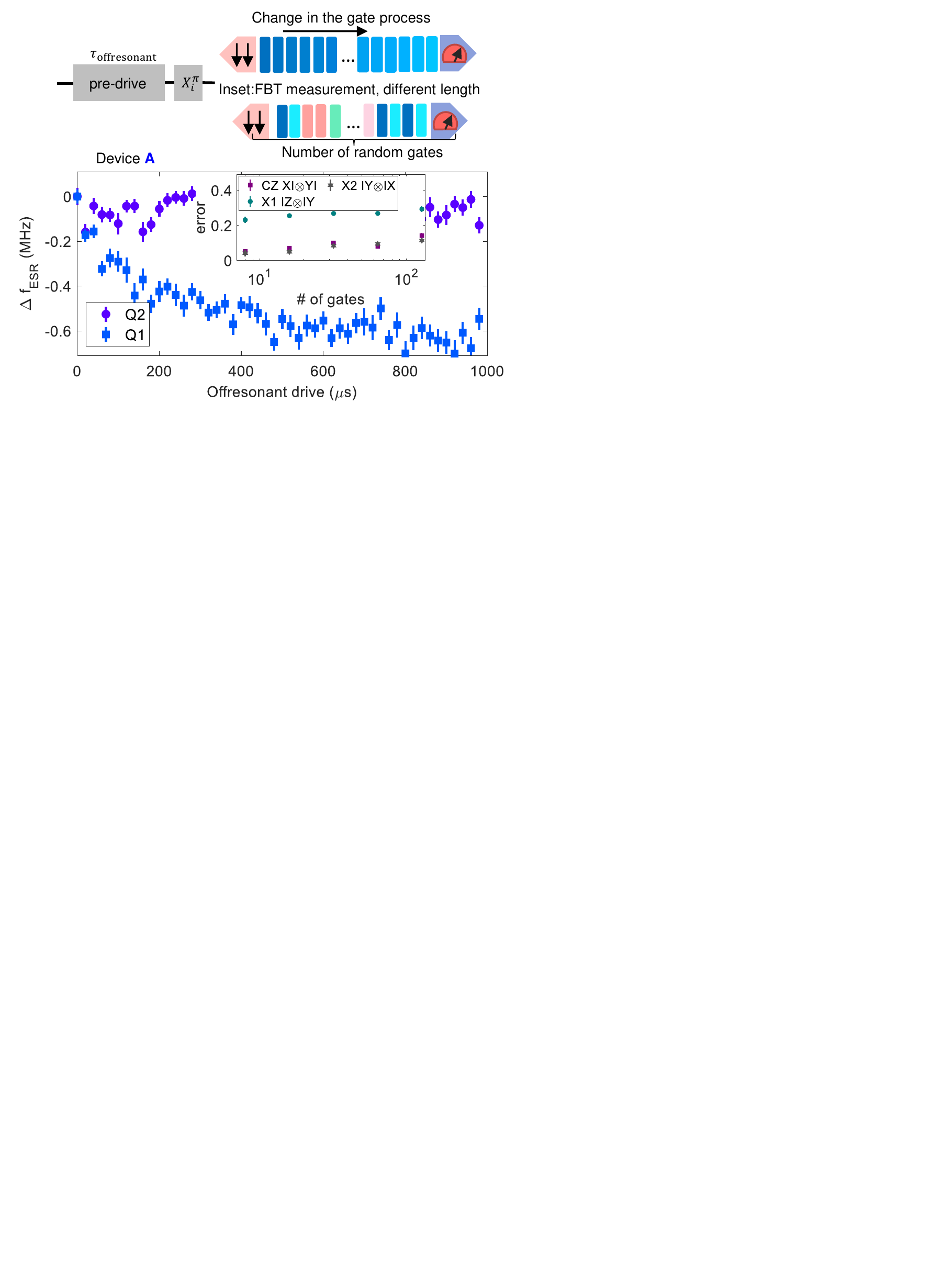}
    \caption{\textbf{Contextual Larmor frequency shift and errors.} 
    The Larmor frequency of both qubits in device A after applying an offresonant pre-pulse. The Q1 frequency shifts to lower frequencies with longer applied pre-pulse, due to a transient effect, possibly due to transient effects similar to Refs.~\cite{Freer_2017_singletom,Takeda2018optimized,undseth2023hotter}. This leads to contextual errors. \textbf{Inset}, Estimated gate noise channel components XI~$\otimes$~YI for CZ, IZ~$\otimes$~IY for $X_1^{\pi/2}$ and IY~$\otimes$~IX for $X_2^{\pi/2}$ as a function of number of gates in a sequence. The shift indicates the changes in the process over the time of one shot of experiment. 
}
    \label{fig:contextual}
\end{figure}

We note that some errors such as the ones in Fig.~\ref{fig:MajorErrors}g remain unexplained, such as the strong presence of IY Hamiltonian errors even in the CZ implementation (which does not contain any microwave pulses) and the appearance of a stochastic XZ error in both devices. These errors are consistent enough across devices and measurement setups that they instigate future investigations of yet unknown microscopic physical mechanisms impacting spins in quantum dots.~\color{black} Together, these sources of error add to a significant amount and are part of the impediment to achieving the next level of gate fidelities.~\color{black}

Another striking observation that we obtain both from GST and FBT in extended Table~\ref{extTable:Fidelity} is that for all devices the single qubit gates are the lowest in fidelity. A more careful GST analysis, in extended Table~\ref{tab:generator_infdl_devB}, reveals that on-target fidelity -- that is, the fidelity of the operation on one qubit discarding effects on the other qubit -- is very high. The idling qubit, however, suffers strong impacts from both dephasing while idling and crosstalk considered in the total error.

\color{black}A caveat for all of our analysis methods is that they all intrinsically assume that the gates and their associated errors are Markovian processes. We have identified physical effects that do not agree with this assumption~\cite{su2023characterizing}. Fig.~\ref{fig:contextual}, for instance, shows that the qubit Larmor frequencies shift significantly depending on how long the microwave has been operating -- an effect potentially associated to the excitation of the two-level fluctuators in the oxide~\cite{undseth2023hotter}. This leads to consistent biases in calibration for the gates applied late in a long circuit compared to the same gate applied early -- a form of contextual noise~\cite{Takeda2018optimized} which is shown in the inset of Fig.~\ref{fig:contextual}. Another non-Markovian effect is the high-amplitude, low-frequency components in noise stemming from either nuclear spins or the slow two-level fluctuators that compose the $1/f$ electric noise spectrum~\cite{Chan2018assessment}. In these circumstances it becomes evident that the differences in the statistical inference approach (frequentist or Bayesian) and the length of circuits (shorter circuits in GST or longer random Cliffords for IRB and FBT) will provide different information about the gates and how they change over time. We also note that depending on the statistical treatment of the data, these contextual drifts in Hamiltonian error would be interpreted as stochastic noise.  
\color{black}
\color{black}
\subsection{Two-Qubit Fidelity Statistics}

\begin{figure}
    \centering
    \includegraphics[width = 1.5\columnwidth, angle = 0,trim = 0cm 16.8cm 2.5cm 0cm]{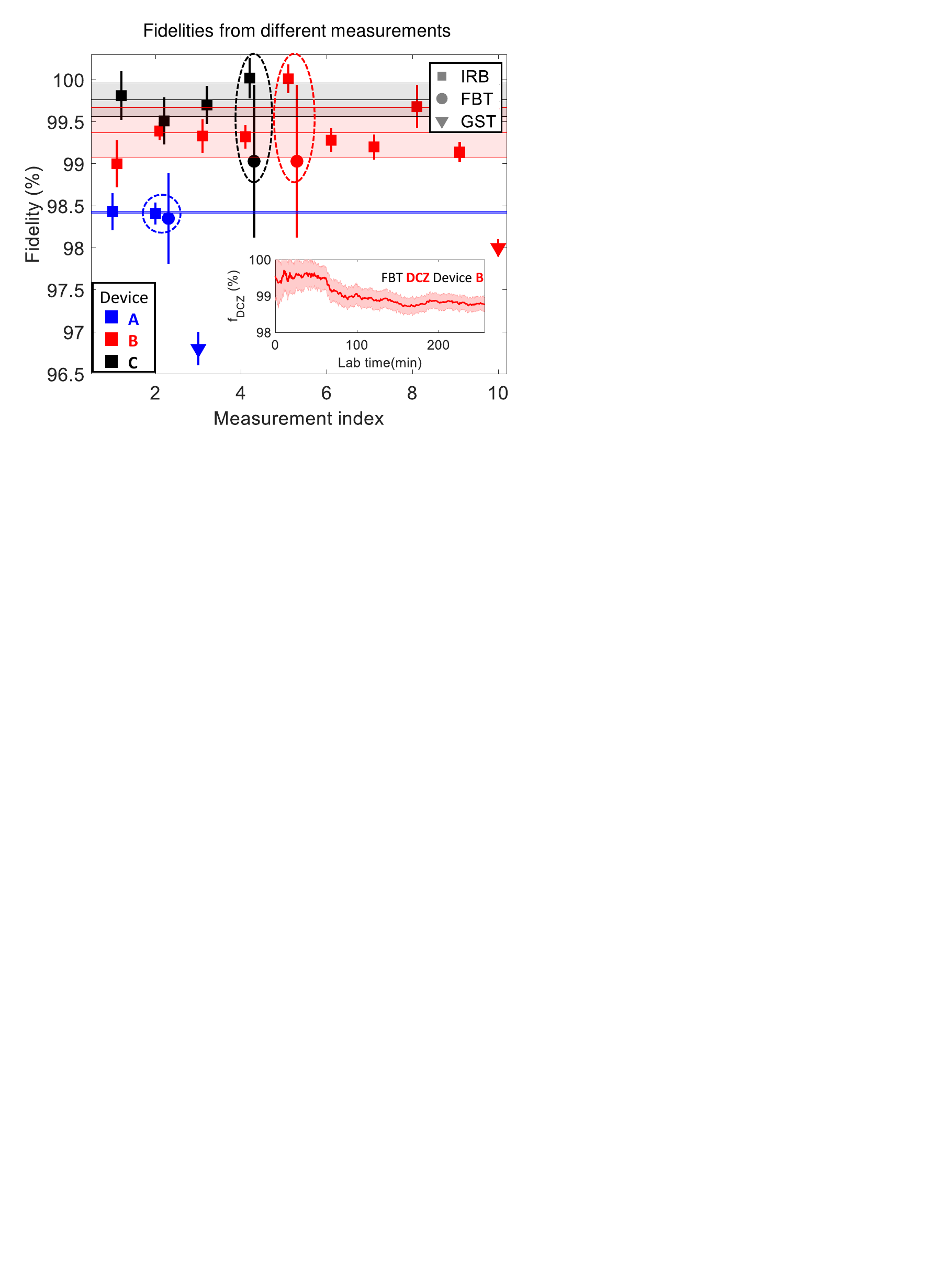}
    \caption{\color{black}\textbf{Two qubit gate fidelities.} Two-qubit gate fidelities extracted from different error characterization measurements of IRB, and GST. The fidelity of the FBT analysis from certain IRB is the neighbouring data point. This data is also shown in the Extended Data Table~\ref{extTable:Fidelity}. Error bars are the errors of the fit. The average and the standard deviation of the IRB measurements is indicated with lines and shaded areas around the line. Each IRB run consists of 200 (A) or 500 (B \& C) circuit randomizations. Inset, DCZ fidelity as a function of lab time, where the transient fidelities are extracted from FBT analysis on IRB experiment data. The shaded region indicates the 95\% confidence interval of the fidelity based on sampling.\color{black}}\label{fig:Fidelity}
\end{figure}

Achieving a detailed picture of errors such as the one in Fig.~\ref{fig:MajorErrors} requires a minimum level of stability of high fidelity operations such that long experiments can be run with consistent results over long periods of time. Our evaluations of two-qubit fidelities from different experimental runs are shown in Fig.~\ref{fig:Fidelity} as well as in the Extended data Table~\ref{extTable:Fidelity}. Most statistics is collected using the simplest validation method -- IRB. Based on IRB experiments, we achieved average two-qubit entangling gate fidelities of 98.4\% (CZ), 99.37\% (DCZ), and 99.78\% (DCZ) in devices A, B, and C respectively. These numbers indicate sufficient operational fidelity for sustainable error correction~\cite{fowler2012surface, Stace2009threshold, Stace2010errorcorrection, Auger2017Fault}. More details on the IRB method can be found in Supplementary Discussion: Randomised benchmarking.

We note, however, that non-Markovian effects create significant challenges to the IRB approach and can lead to an overestimated fidelity~\cite{Fogarty2015nonexponential}. In some instances, we found that the circuits performed better when interleaved with DCZ gates, resulting in unphysical estimated fidelities surpassing 100\%. On the other hand, relying only on GST experiments exposes our results to instability in the active feedback procedure since they are  based on long experiments and a frequentist approach to the statistical treatment of the data. This degradation of the quality of the gates over time is to some extent captured in a Bayesian analysis, as seen by plotting the evolution of the FBT estimates for two-qubit gates during the operation for more than four consecutive hours as seen in the inset of Fig.~\ref{fig:Fidelity}.

The FBT analysis is performed directly on the outcomes of any arbitrary input circuit (we note that GST may also be adapted for analysing arbitrary circuits). We utilise the data from IRB runs as input, decomposing all the circuits into their five primitive gates (for additional information see also Supplementary Discussion: Implementation of fast Bayesian tomography).

To understand better the experimental runs yielding unphysical fidelities (>100\% in B and C) we chose those for the FBT analysis. Tomographic analysis that yields a physical process matrix cannot have above 100\% fidelity and FBT has the additional benefit of allowing the analysis of the process in lab time through Bayesian inference~\cite{Evans2022FastBayesian}. The fidelities estimated across the full experiment by FBT in Fig.~\ref{fig:Fidelity} show a very large uncertainty, which can be understood from the plot of the estimated fidelity of the DCZ gate in device B as a function of lab time. Over time, the gate calibration degrades and infidelities more than double.

\color{black}

\subsection{Implications to large scale spin quantum computers}

Besides the benefits for operation at scale, the consistency of the operations demonstrated here significantly improves the quality of the physical conclusions that can be drawn from experiments. We were able to identify commonalities between gates implemented in different devices, setups and adopting different strategies.

The stability of the operations over time, combined with the theoretical and methodological improvements in process tomography, opens a window into the physics of spin qubit errors.~\color{black} With the possibility of consistently performing high-fidelity one- and two-qubit gates for several hours, we can form hypotheses and test them in a repeatable manner over many hours or days of experiments.~\color{black} Our work not only offers compelling evidence for the microscopic nature of some of the error sources, but it also reveals some unknown processes with, as yet, unclear physical origins.

One of the most important conclusions is that the entangling gate fidelities achieved here can be systematically improved with a combination of better materials to reduce noise~\cite{elsayed2022low,Wan2023Foundry}, active use of these tomographic results to recalibrate gates against Hamiltonian errors~\cite{xue2022quantum,su2023characterizing} and pulse engineering to reduce the stochastic errors with robust gates and dynamical decoupling~\cite{Gungord2019Analytically, Gungord2022Robust, Tang2023Designing,yang2019silicon}~\color{black} (see also: Supplementary Discussion: Comparison with other high fidelity two spin systems in silicon). The fidelity estimates quoted here are far from being fundamentally limited by the physics of spin qubits or by the minimum noise levels in materials. ~\color{black}

The scalability of the traditional Loss and DiVincenzo approach\cite{LossDiVincenzo1998} to spin qubits can be obscured by the overhead imposed by the strategies required for high-fidelity operation. Due to the relative phases, the number of parameters requiring feedback grows prohibitively fast with the number of qubits if no mitigation strategy is adopted~\cite{Philips20226qubit}. Moreover, circumventing errors with increasingly convoluted engineered pulses tailored to the idiosyncrasies of each qubit leads to control signals that are hard to generate at scale in an automatised manner. Finally, the degraded performance of idling qubits leads to a steep price in multiqubit operation. Innovative control solutions might be required to evade these difficulties, including continuously dynamically-decoupled driven qubit implementations and pulse shapes aimed at maximising performance regardless of the particular properties of each qubit\cite{Vahapoglu2021dielectric, Hansen2021PulseEngineering, Hansen2022Implementation, Seedhouse2021Quantum}. 

Spin qubits in MOS-based quantum dots now join the select group of qubit technologies with two-qubit gate fidelities exceeding~99\% barrier --- measured average IRB fidelity being 99.17\% for three devices with standard deviation of 0.56\%. The prospects for fault tolerant operation with this platform are further enhanced when the structure of the errors is taken into account -- the strong bias towards dephasing errors instead of depolarising errors opens up the possibility for significant gains in error correction code performance. This will help push the average fidelity up and standard deviation down.

\section{Methods}

\subsection{Experimental devices}

The three devices studied in this work were fabricated using multi-level aluminium gate-stack silicon MOS technology~\cite{angus2008silicon,lim2009observation} on isotopically enriched silicon-28 substrates of~800~ppm residual \textsuperscript{29}Si (Device A \& B) and of~50~ppm residual \textsuperscript{29}Si (Device C). A layer of SiO$_2$ of $\sim$~8~nm was thermally grown above the silicon substrates. The devices are designed with plunger gate width of~30~nm and gate pitch as small as~50~nm. This allows a~20~nm gap between the plunger gates for J-gate. 

\subsection{Measurement Setup}

Device A was measured in an Oxford Kelvinox 400HA dilution refrigerator. DC bias voltages were generated from Stanford Research Systems SIM928 Isolated Voltage Sources. Gate pulse waveforms were generated by a Quantum Machines (QM) OPX+ and combined with DC biases using custom linear bias combiners at room temperature.

Devices B \& C were measured in a Bluefors XLD400 dilution refrigerator. DC bias voltages were generated with Basel Precision Instruments SP927 DACs. Gate pulse waveforms were generated by a Quantum Machines OPX and combined with DC biases using custom linear bias combiners at the~4~K stage.

The SET current of device A was amplified using a room temperature I/V converter (Basel SP983c) and sampled by a QM OPX. The SET of devices B \& C were connected to a tank circuit for reflectometry measurement, with the tone  generated by the QM OPX. The return signal was amplified by a Cosmic Microwave Technology CITFL1 LNA at the 4~K stage, and a Mini-circuits ZX60-P33ULN+ and a Mini-circuits ZFL-1000LN+ at room temperature, before digitised and demodulated by the QM OPX.

For all devices, microwave pulses were generated with a Keysight PSG8267D Vector Signal Generator, with I/Q and pulse modulation waveforms generated by the QM OPXs.

\section{Data Availability}
The datasets generated and/or analyzed during this study are available from the corresponding authors on reasonable request.

\section{Code Availability}
The analysis codes that support the findings of the study are available from the corresponding authors on reasonable request.

\section{Acknowledgements}
We thank A. Dickie and S. Serrano for technical help, T. Evans for help with FBT, H. Stemp with GST, and A. Noiri with IRB. We acknowledge support from the Australian Research Council (FL190100167 and CE170100012), the US Army Research Office (W911NF-17-1-0198, W911NF-23-10092), and the NSW Node of the Australian National Fabrication Facility. The views and conclusions contained in this document are those of the authors and should not be interpreted as representing the official policies, either expressed or implied, of the Army Research Office or the US Government. The US Government is authorized to reproduce and distribute reprints for Government purposes notwithstanding any copyright notation herein. J.Y.H., W.G., R.S., M.K.F., A.E.S, and J.D.C. acknowledge support from Sydney Quantum Academy. C.O. S.S. K.M.R. and R.B.-K acknowledge support by the U.S. DOE/SC/ASCR through the Quantum Testbed Pathfinder Program. Sandia is a multiprogram laboratory managed and operated by NTESS, LLC., a wholly owned subsidiary of Honeywell International, Inc., for the U.S. DOE’s NNSA under Contract No. DE-NA-0003525. 

\section{Author information}
\subsection{Author Contributions}

W.H.L. and F.E.H. fabricated the devices, with A.S.D.'s supervision, on isotopically enriched $^{28}$Si wafers supplied by K.M.I. (800~ppm), N.V.A., H.-J.P., and M.L.W.T. (50~ppm). 
T.T. and W.H.L. measured initial devices and developed the final J-gate design. T.T. W.H.L. J.Y.H., N.D.S. and W.H.L. did the experiments, coding and initial analysis, with A.M., A.L., A.S., C.H.Y., and A.S.D.'s supervision. 
R.C.C.L and W.H. helped with the initial experimental designs and ideas.
R.Y.S. did the FBT analysis with supervision from S.D.B. and A.S.
M.K.F. together with C.O., S.S., K.M.R. and R.B.-K. did the GST analysis.
A.E.S. and N.D.S. did the feedback analysis.
J.D.C. simulated the device structure to guide the gate design with supervision from C.C.E. and A.S.
T.T. wrote the manuscript, with the input from all authors.

\subsection{Corresponding Authors}
Correspondence to the first or last authors.

\section{Competing Interests}
A.S.D. is CEO and a director of Diraq Pty Ltd. Other authors declare no competing interest.

\bibliographystyle{naturemag}
\bibliography{2Qfidelitybibcleaned}

\ifsupp

\makeatletter
\renewcommand{\figurename}{Extended Data Figure}
\makeatother

\newcolumntype{K}{>{\vspace*{0.11mm}\centering}X}
\newcolumntype{C}{>{\vspace*{0.11mm}\centering}c}

\renewcommand{\tablename}{Extended Data Table}
\makeatletter
\renewcommand*{\fnum@table}{{\normalfont\bfseries \tablename~\thetable}}
\makeatother

\captionsetup[table]{name={Extended Data Table},labelsep=period,justification=centerlast,font=small}
\captionsetup[figure]{name={\bf{Extended Data Fig.}},labelsep=line,justification=centerlast,font=small}
\setcounter{figure}{0}

\setcounter{table}{0}

\begin{table}[htb]
\caption{Error characterization methods used in this paper.}\label{ExtTable:QCVVmethods}
\resizebox{\columnwidth}{!}{%
\begin{tabular}{|c|c|c|c|}
\cline{1-4}
Method& Measurement &Analysis & process matrix  \cr  \hline\hline
IRB & Randomised Cliffords & decay fit & no \cr  \cline{1-4} %
FBT& Any gate sequence& Bayesian estimation & yes \cr  \cline{1-4} %
GST&Periodic info-complete fiducials & Maximum-likelyhood& yes \cr  \cline{1-4} %
\end{tabular}
}
\end{table}

\begin{table*}[t]
\centering
\caption{\textbf{Device parameters.} Comparison table of the characteristics and operation principles of all the devices used in this experimental campaign.}\label{extTable:DeviceComparison}
\begin{tabular}{|c|c|c|c|c|c|c|l}
\cline{1-7}
 Device & \multicolumn{2}{c|}{Device \color{blue}\textbf{A} \color{black}} & \multicolumn{2}{c|}{Device \color{red}\textbf{B} \color{black} } & \multicolumn{2}{c|}{Device \color{black}\textbf{C} \color{black} }& \\ \cline{1-7}
 $T_{2} ^*$($\mu$s) Q1,Q2& $5.5 \pm 0.3$&$6.5 \pm 0.3$ & $3.0 \pm 0.3$ & $2.6 \pm 0.3$  & $5.5 \pm 0.3$ &$5.5 \pm 0.3$ &  \\ \cline{1-7}
 $T_{2} ^{\textrm{Hahn}}$($\mu$s) Q1,Q2& $51 \pm 3$ &$60 \pm 3$&$39 \pm 3$ & $46 \pm 3$  &$132 \pm 3$ & $72 \pm 3$ &  \\ \cline{1-7}
 $T_{2} ^{\textrm{Rabi}}$($\mu$s) Q1,Q2& $55 \pm 5$ &$64 \pm 5$&$26 \pm 3$ & $56 \pm 3$  & $131 \pm 5$ & $100 \pm 5$  &  \\ \cline{1-7}
 $X ^{\pi/2}$($\mu$s) Q1,Q2& $ 0.520 $ &$0.400$&$0.200$ & $0.380$  & $0.326$ & $0.200$  &  \\ \cline{1-7}
 1Q RB fidelity& 99.5\%&99.6\% & 99.6\%&99.7\% & - & - &  \\ \cline{1-7}
 $\Delta E_Z$ (MHz)& \multicolumn{2}{c|}{10.8} & \multicolumn{2}{c|}{22.7} & \multicolumn{2}{c|}{12.95}& \\ \cline{1-7}
 J-gate turn on (dec/V)& \multicolumn{2}{c|}{17.13} & \multicolumn{2}{c|}{45.15} & \multicolumn{2}{c|}{20.64}& \\ \cline{1-7}
 J-gate swing (mV)& \multicolumn{2}{c|}{330.4} & \multicolumn{2}{c|}{72.24} & \multicolumn{2}{c|}{52.48}& \\ \cline{1-7}
 Exchange time CZ/DCZ (ns)& \multicolumn{2}{c|}{256} & \multicolumn{2}{c|}{500} & \multicolumn{2}{c|}{990}& \\ \cline{1-7}
 Exchange on (kHz)& \multicolumn{2}{c|}{1950} & \multicolumn{2}{c|}{1000} & \multicolumn{2}{c|}{505}& \\ \cline{1-7}
 Exchange off (Hz)& \multicolumn{2}{c|}{4.2} & \multicolumn{2}{c|}{550} & \multicolumn{2}{c|}{<13700}& \\ \cline{1-7}
 Initialization & \multicolumn{2}{c|}{$T^-$ (Heralded)} & \multicolumn{2}{c|}{$\uparrow\downarrow$} & \multicolumn{2}{c|}{$T^-$ (Heralded)}& \\ \cline{1-7}
 $B_0$ and orientation& \multicolumn{2}{c|}{0.8274 T, [100]} & \multicolumn{2}{c|}{0.893 T, [110]} & \multicolumn{2}{c|}{0.790 T, [110]}& \\ \cline{1-7}
  Charge configuration & \multicolumn{2}{c|}{(1,3) Isolated} & \multicolumn{2}{c|}{(1,3) Reservoir} & \multicolumn{2}{c|}{(3,3) Reservoir}& \\ \cline{1-7}
  Parity readout visibility & \multicolumn{2}{c|}{98 \%} & \multicolumn{2}{c|}{99 \%} & \multicolumn{2}{c|}{95 \%}& \\ \cline{1-7}
  GST gauge \textit{invariant} SPAM fidelity & \multicolumn{2}{c|}{97.6$\pm$0.9 \%} & \multicolumn{2}{c|}{98.4$\pm$0.4 \%} & \multicolumn{2}{c|}{-}& \\ \cline{1-7}
  Single shot time ($\mu$s) & \multicolumn{2}{c|}{1000} & \multicolumn{2}{c|}{100} & \multicolumn{2}{c|}{200}& \\ \cline{1-7}
  Residual $^{29}$Si & \multicolumn{2}{c|}{800 ppm} & \multicolumn{2}{c|}{800 ppm} & \multicolumn{2}{c|}{50 ppm}& \\ \cline{1-7}
  Design \# dots & \multicolumn{2}{c|}{three} & \multicolumn{2}{c|}{three} & \multicolumn{2}{c|}{four}& \\ \cline{1-7}
  Feedbacks & \multicolumn{2}{c|}{$I_\textrm{SET}$,$\epsilon$,$f_\textrm{ESR}^{\textrm{Qi}}$,$P_{Xi}^{\textrm{Qi}}$,$V_\textrm{CZ}$,$\varphi_\textrm{CZ}^\textrm{Qi}$} & \multicolumn{2}{c|}{$I_\textrm{SET}$,$\epsilon$,$f_\textrm{ESR}^{\textrm{Qi}}$,$P_{Xi}^{\textrm{Qi}}$,$V_\textrm{DCZ}$} & \multicolumn{2}{c|}{$I_\textrm{SET}$}& \\ \cline{1-7}
\end{tabular}
\end{table*}

\begin{table*}[t]
\centering
\caption{Fidelities extracted from several experimental error characterization runs. FBT results are from one of the IRB runs indicated by italics. Mean and standard deviation (std) are calculated from all IRB runs for a given device. The error bars for IRB come from the error bars of the fit and for FBT we report the average fidelity over the experiment with error bars being the 2$\sigma$ variation over the experiment.~\color{black}Clifford fidelity is the average fidelity of a uniformly random 2-qubit Clifford operation. two-qubit Clifford operations are compiled into native gates, and each one includes (on average) 1.5 CZ/DCZ gates and 1.8 single-qubit  $X_\frac{\pi}{2}$ gates on each of the two qubits.\color{black} For GST the error bars are given by the error of the log-likelyhood fit. Some of these are also plotted in the main text Figure~\ref{fig:Fidelity}.}
\label{extTable:Fidelity}
\begin{tabular}{|c|c|c|c|c|}
\cline{1-5}
Run/Gate  & Clifford (\%) & \color{violet}CZ\color{black}/\color{darkgray}DCZ\color{black} (\%) & X$^{\pi/2}_1$ (\%) & X$^{\pi/2}_2$ (\%) \cr  \cline{1-5}
IRB$_{1}$ \color{blue} \textbf{A} \color{black} & $90.78\pm 0.10$ & \color{violet}$98.43\pm 0.22$\color{black} & - & - \cr  \cline{1-5} %
\emph{IRB}$_{2}$ \color{blue} \textbf{A} \color{black} & $90.71\pm 0.06$ & \color{violet}$98.41\pm 0.13$\color{black} & - & - \cr  \cline{1-5} %
mean,std\color{blue} \textbf{A} \color{black} & $90.75\pm 0.05$ & \color{violet}$98.42\pm 0.01$\color{black} & - & - \cr  \cline{1-5} %
IRB$_{3}$ \color{red}\textbf{B} \color{black}  & $86.95\pm 0.19$ & \color{darkgray}$99.00\pm 0.28$\color{black} & - & - \cr  \cline{1-5} 
IRB$_{4}$ \color{red}\textbf{B}  \color{black}  & $89.65\pm 0.06$ & \color{darkgray}$99.39\pm 0.11$\color{black} & - & - \cr  \cline{1-5} 
IRB$_{5}$ \color{red}\textbf{B}  \color{black}  & $85.06\pm 0.10$ & \color{darkgray}$99.33\pm 0.20$\color{black} & - & - \cr  \cline{1-5} 
\emph{IRB}$_{6}$ \color{red}\textbf{B}  \color{black}  & $84.93\pm 0.09$ & \color{darkgray}$100.01\pm 0.17$\color{black} & - & - \cr  \cline{1-5} 
IRB$_{7}$ \color{red}\textbf{B}  \color{black}  & $86.13\pm 0.07$ & \color{darkgray}$99.32\pm 0.14$\color{black} & - & - \cr  \cline{1-5} 
IRB$_{8}$ \color{red}\textbf{B}  \color{black}  & $87.52\pm 0.07$ & \color{darkgray}$99.28\pm 0.14$\color{black} & - & - \cr  \cline{1-5} 
IRB$_{9}$ \color{red}\textbf{B}  \color{black}  & $87.29\pm 0.08$ & \color{darkgray}$99.20\pm 0.15$\color{black} & - & - \cr  \cline{1-5} 
IRB$_{10}$ \color{red}\textbf{B}  \color{black}  & $86.90\pm 0.11$ & \color{darkgray}$99.68\pm 0.26$\color{black} & - & - \cr  \cline{1-5} 
IRB$_{11}$ \color{red}\textbf{B}  \color{black}  & $87.19\pm 0.07$ & \color{darkgray}$99.14\pm 0.12$\color{black} & - & - \cr  \cline{1-5} 
mean,std \color{red}\textbf{B}  \color{black}  & $86.85 \pm 1.41$ & \color{darkgray}$99.37\pm 0.3$\color{black} & - & - \cr  \cline{1-5} %
IRB$_{12}$ \color{black}\textbf{C} \color{black}  & $89.69\pm 0.11$ & \color{darkgray}$99.81\pm 0.29$\color{black} & - & - \cr  \cline{1-5} 
IRB$_{13}$ \color{black}\textbf{C}  \color{black}  & $88.02\pm 0.13$ & \color{darkgray}$99.51\pm 0.28$\color{black} & - & - \cr  \cline{1-5} 
IRB$_{14}$ \color{black}\textbf{C} \color{black}  & $90.52\pm 0.09$ & \color{darkgray}$99.70\pm 0.23$\color{black} & - & - \cr  \cline{1-5} 
\emph{IRB}$_{15}$ \color{black}\textbf{C}  \color{black}  & $89.08\pm 0.10$ & \color{darkgray}$100.02\pm 0.24$\color{black} & - & - \cr  \cline{1-5} 
mean,std \color{black}\textbf{C} \color{black}  & $89.33\pm 1.05$ & \color{darkgray}$99.76\pm 0.20$\color{black} & - & - \cr  \cline{1-5} 
FBT \color{blue} \textbf{A}  \color{black}  & - & \color{violet}$98.35\pm 0.54$\color{black} & $98.41\pm 0.89$&$99.09\pm 0.78$ \cr  \cline{1-5}
FBT \color{red}\textbf{B}  \color{black}  & - & \color{darkgray}$99.03\pm 0.91$\color{black} & $95.92\pm 0.78$&$96.58\pm 0.66$ \cr  \cline{1-5}
FBT \color{black} \textbf{C}  \color{black}  & - & \color{darkgray}$99.04\pm 0.54$\color{black} & $97.00\pm 1.16$&$97.72\pm 0.87$ \cr  \cline{1-5}
GST \color{blue} \textbf{A} \color{black}  & - & \color{violet}$96.8\pm 0.2$\color{black} & $96.5\pm 0.2$&$97.5\pm 0.2$ \cr  \cline{1-5}
GST \color{red}\textbf{B}  \color{black}  & - & \color{darkgray}$98.0\pm 0.1$\color{black} & $97.9\pm 0.2$&$95.7\pm 0.2$ \cr  \cline{1-5}
\end{tabular}
\end{table*}

\begin{figure*}[ht!]
    \centering
    \includegraphics[width = 1\textwidth, angle = 0, trim = 0cm 20cm 0cm 0cm]{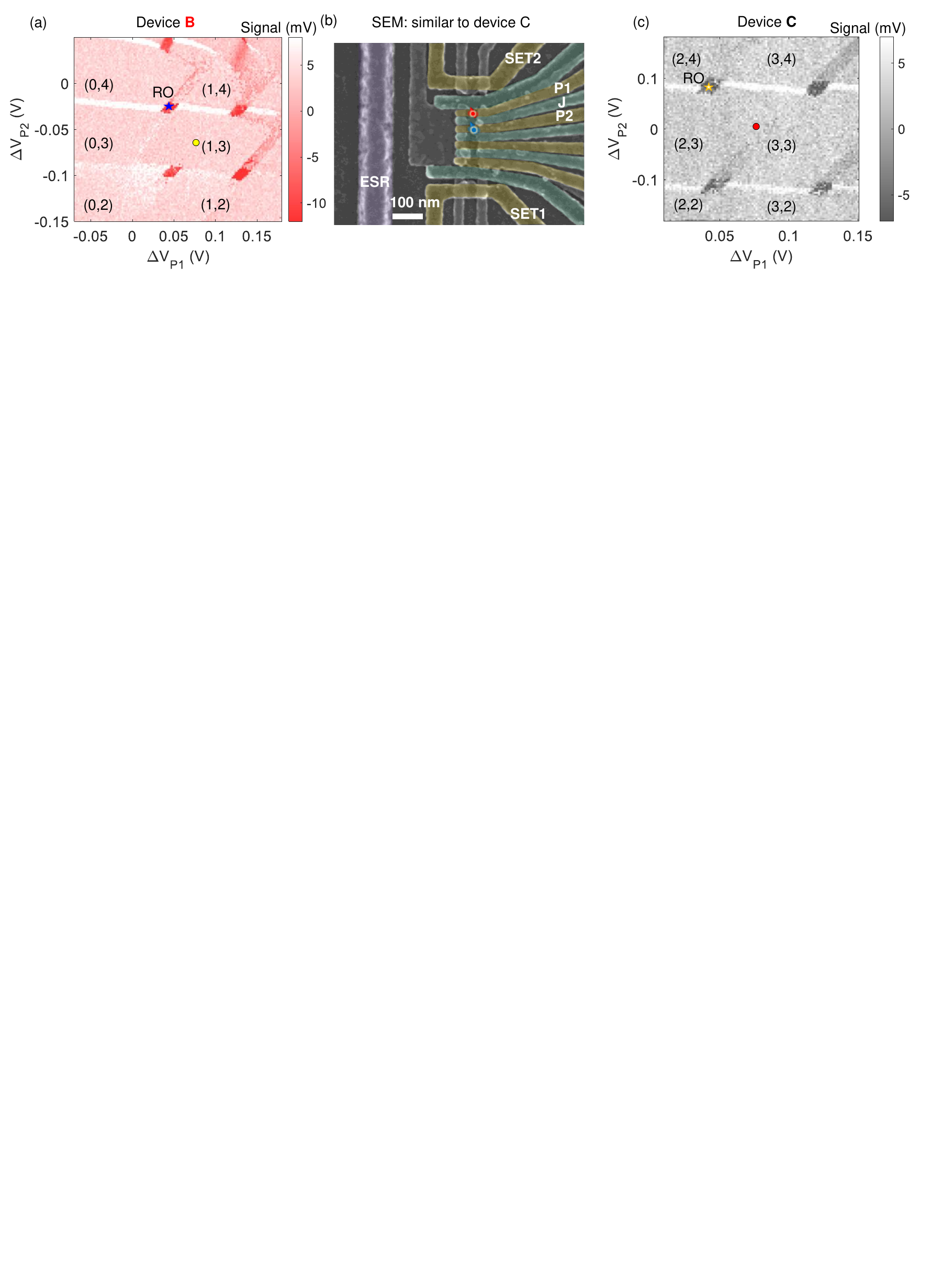}
    \caption{\textbf{Device B and C stability maps and SEM of a device similar to C.}
    \textbf{a}, Device B stability map. Operation point (coloured circle) and readout point (RO) are indicated, together with the charge occupancies. 
    \textbf{b}, Scanning electron micrograph (SEM) of a device similar to C.
    \textbf{c}, Device C stability map. Indication same as in a.}
    \label{extFig:StabilityMaps}
\end{figure*}


\begin{figure*}[ht!]
    \centering
    \includegraphics[width = 0.8\textwidth, angle = 0, trim = 0cm 9cm 0cm 1cm]{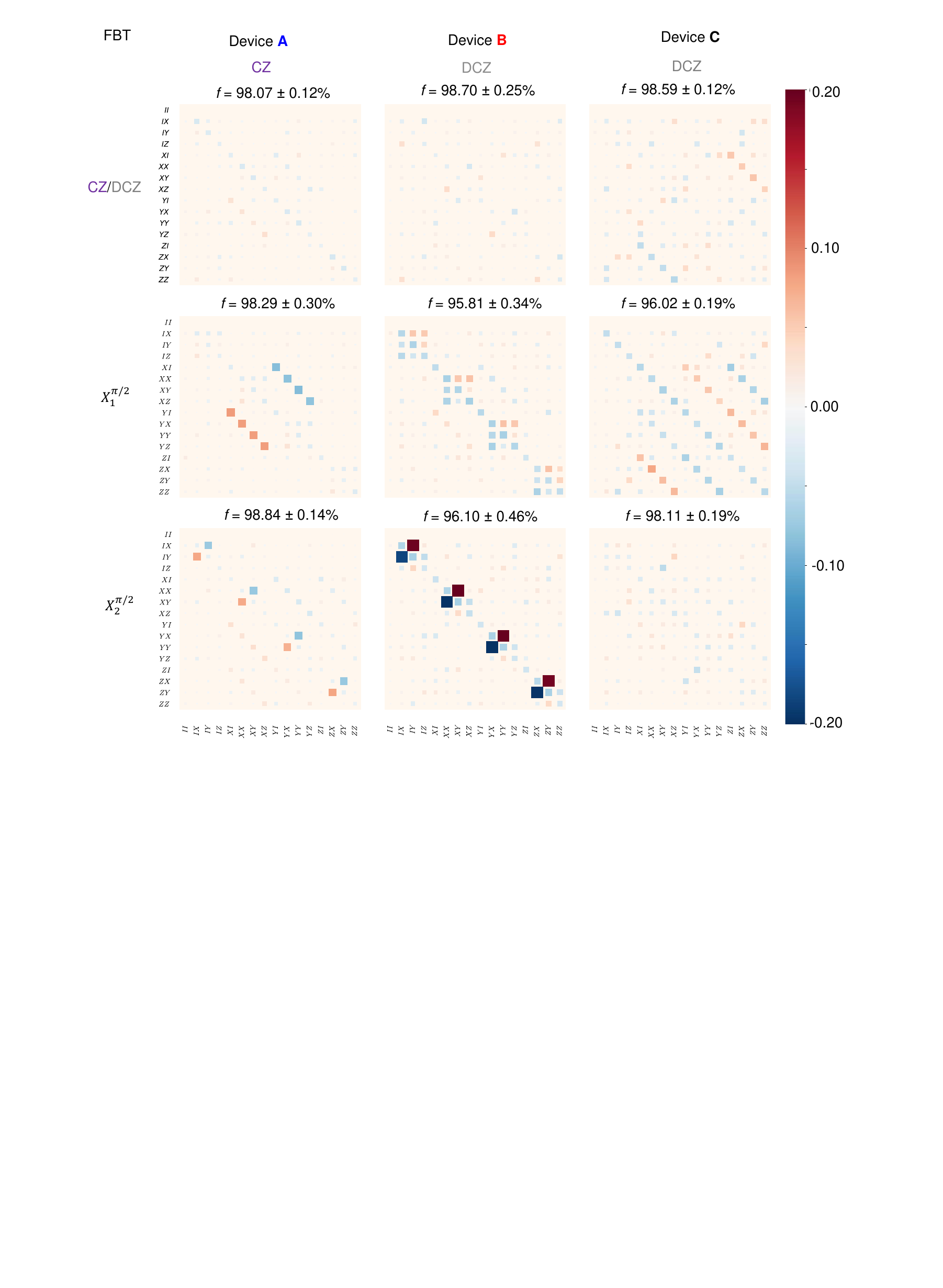}
    \caption{\textbf{Error generators from FBT analysis for all three devices, for single and two qubit gates.}
    FBT estimated gate error generators for each device are analyzed from IRB experiments individually. Plotted error generators are the posterior of the last update of each analysis. Fidelity and its uncertainty (with $95\%$ confidence interval) are calculated by re-sampling with the Gaussian statistics of the final estimated results.}  
    \label{extFig:FBTerrorgenerators}
\end{figure*}

\begin{table*}
	\centering
	\caption{Error metrics from GST characterization of (a) device A and (b) device B, partitioned by qubit support. Generator infidelities and total error are defined in equations 4 and 5 of the supplementary material. For both metrics, we have broken down the total infidelity/error into their contributions from each of the qubits. The on-target generator fidelity corresponds to the fidelity with which a given gate enacts the ideal gate operation on the qubit being targeted and the total error takes into account the errors occurring in the spectator qubit while idling.\color{black} By breaking down the infidelity based on the support of the operations we can see that a significant contribution to the single qubit gate error budget comes from crosstalk errors inducing undesired evolution on the spectator qubit.}
	\begin{subtable}[c]{\textwidth}
	\centering
	\begin{tabular}{|c|c|c|c|c|c|c|c|}
		\hline
		\multirow{2}[4]{*}{Gate} & \multicolumn{1}{c|}{\multirow{2}[4]{*}{\shortstack{On-Target \\ Generator Fidelity}}} & \multicolumn{3}{c|}{Generator Infidelity} & \multicolumn{3}{c|}{Total Error} \bigstrut\\
		\cline{3-8}          &       & On Q1 & On Q2 & Total & On Q1 & On Q2 & Total \bigstrut\\
		\hline
		$X_1^{\pi/2}$ & 99.3(4)\% & \textbf{0.7(4)\%} & 3.5(4)\% & 4.5(8)\% & 7.6(5)\% & 9.8(6)\% & 13.6(8)\% \bigstrut\\
		\hline
		$Z_1^{\pi/2}$ & 99.9(3)\% & \textbf{0.1(3)\%} & 0.2(3)\% & 0.3(6)\% & 0.3(4)\% & 2.5(4)\% & 2.8(7)\% \bigstrut\\
		\hline
		$X_2^{\pi/2}$ & 99.1(4)\% & 2.0(4)\% & \textbf{0.9(4)\%} & 3.3(7)\% & 7.0(5)\% & 8.2(4)\% & 11.9(7)\% \bigstrut\\
		\hline
		$Z_2^{\pi/2}$ & 99.9(3)\% & 0.2(3)\% & \textbf{0.1(3)\%} & 0.4(6)\% & 1.1(6)\% & 2.6(4)\% & 3.1(6)\% \bigstrut\\
		\hline
		CZ    & 95.9(7)\% & 1.7(4)\% & 1.5(4)\% & \textbf{4.1(8)\%} & 7.8(4)\% & 7.7(4)\% & 12.9(8)\% \bigstrut\\
		\hline
	\end{tabular}%
    \end{subtable}
    \caption*{(a) Device A}
	\label{tab:generator_infdl_devA}
	\vspace{1em}
	\begin{subtable}[c]{\textwidth}
	\centering
	\begin{tabular}{|c|c|c|c|c|c|c|c|}
		\hline
		\multirow{2}[4]{*}{Gate} & \multicolumn{1}{c|}{\multirow{2}[4]{*}{\shortstack{On-Target \\ Generator Fidelity}}} & \multicolumn{3}{c|}{Generator Infidelity} & \multicolumn{3}{c|}{Total Error} \bigstrut\\
		\cline{3-8}          &       & On Q1 & On Q2 & Total & On Q1 & On Q2 & Total \bigstrut\\
		\hline
		$X_1^{\pi/2}$ & 99.9(3)\% & \textbf{0.1(3)\%} & 2.3(3)\% & 2.7(6)\% & 1.2(5)\% & 4.0(4)\% & 4.7(6)\% \bigstrut\\
		\hline
		$Z_1^{\pi/2}$ & 99.9(6)\% & \textbf{0.1(3)\%} & 0.2(3)\% & 0.4(6)\% & 0.5(4)\% & 1.1(4)\% & 1.4(7)\% \bigstrut\\
		\hline
		$X_2^{\pi/2}$ & 99.9(2)\% & 5.1(3)\% & \textbf{0.1(2)\%} & 5.6(4)\% & 6.4(4)\% & 1.2(4)\% & 7.5(5)\% \bigstrut\\
		\hline
		$Z_2^{\pi/2}$ & 99.9(2)\% & 0.1(2)\% & \textbf{0.1(2)\%} & 0.4(5)\% & 0.4(3)\% & 1.1(3)\% & 1.5(5)\% \bigstrut\\
		\hline
		DCZ   & 97.5(5)\% & 0.7(2)\% & 0.4(2)\% & \textbf{2.5(5)\%} & 1.4(3)\% & 1.3(3)\% & 4.2(6)\% \bigstrut\\
		\hline
	\end{tabular}%
	\end{subtable}
    \caption*{(b) Device B}
	\label{tab:generator_infdl_devB}%
\end{table*}

\begin{figure*}[ht!]
    \centering
    \includegraphics[width = 0.75\textwidth, angle = 0, trim = 0cm 13cm 0cm 0cm]{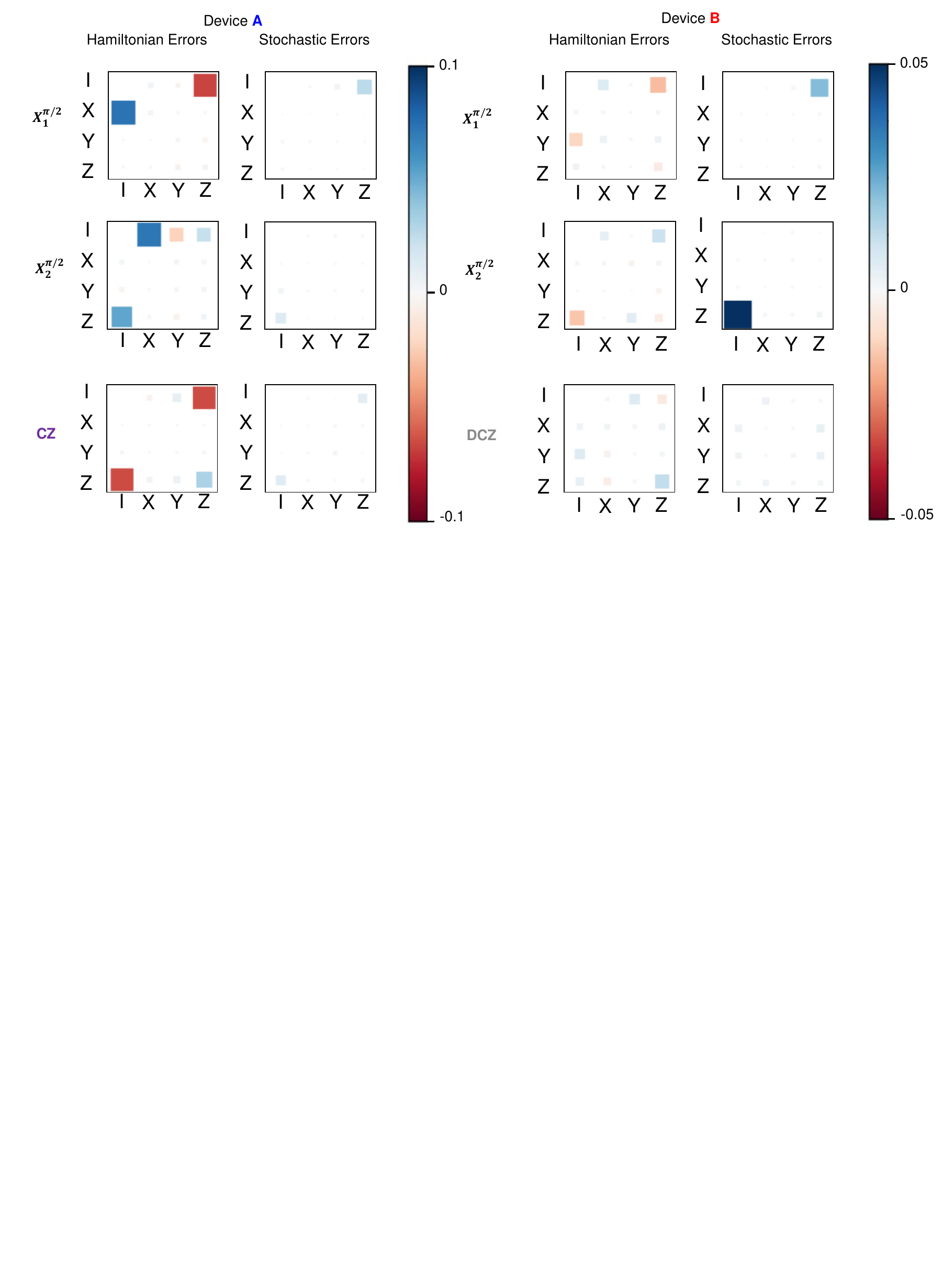}
    \caption{\textbf{Hamiltonian and Stochastic error generators from GST reports for devices A and B.}
    \textbf{a}, Colour bar ranges from -0.1 to +0.1.
    \textbf{b}, Colour bar ranges from -0.05 to +0.05.}
    \label{extFig:GSTerrorDevice}
\end{figure*}

\begin{figure*}[ht!]
    \centering
    \includegraphics[width = 0.8\textwidth, angle = 0, trim = 0cm 0cm 0cm 1cm]{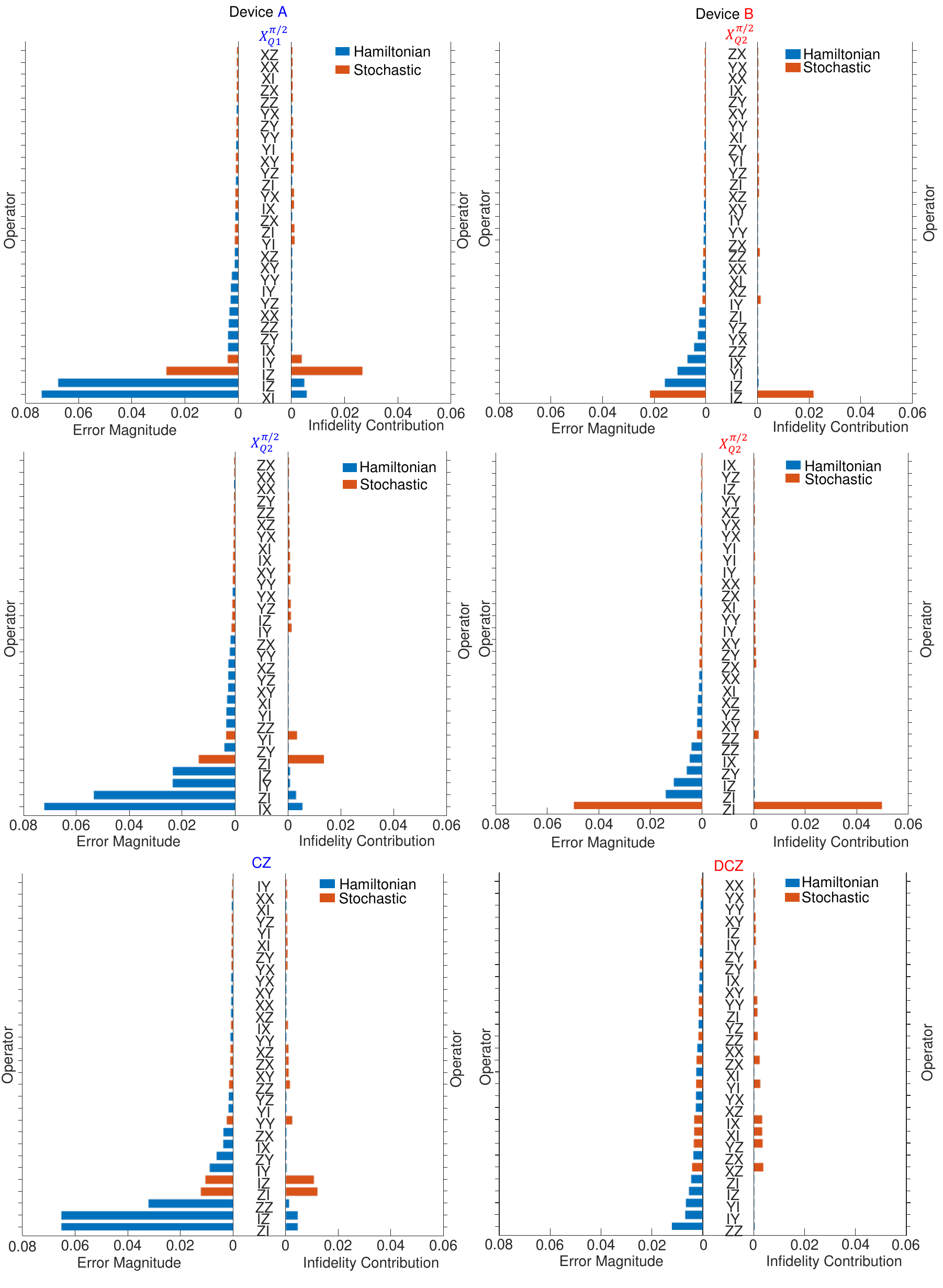}
    \caption{\textbf{Summary of all the error channels GST components.}
    Error channels magnitudes and the infidelity contributions for all the physical gates for devices A and B.~\color{black}The error channels are listed in order of increasing magnitude in the error magnitude. Note that to get the infidelity contribution from error magnitude the Hamiltonian errors are squared, and stochastic errors are the same. A selection of the largest error magnitude components is plotted in the main text in Figs.~\ref{fig:MajorErrors}d, \ref{fig:MajorErrors}e, \ref{fig:MajorErrors}f, and~\ref{fig:MajorErrors}g.\color{black} These are extracted from the process matrices in Extended Figure~\ref{extFig:GSTerrorgenerators}.
   }  
    \label{extFig:GSTallcomponents}
\end{figure*}

\begin{figure*}[ht!]
    \centering
    \includegraphics[width = 0.8\textwidth, angle = 0, trim = 0cm 4cm 0cm 1cm]{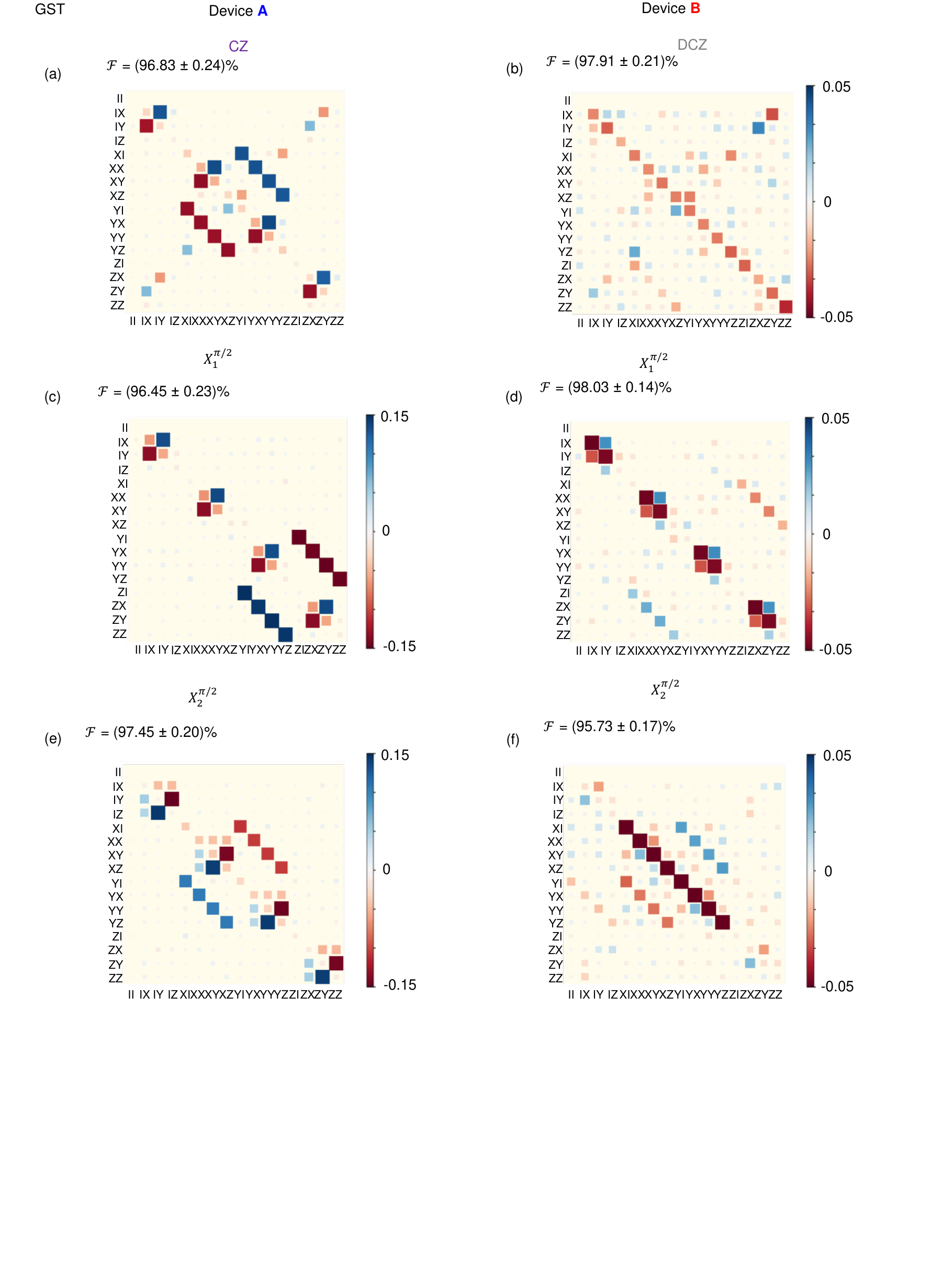}
    \caption{\textbf{Gate Set Tomography, Devices A and B.} 
    \textbf{a}, Full error generator for device A, CZ gate.
    \textbf{b}, Full error generator for device B, DCZ gate.
    \textbf{c}, Full error generator for device A, $\text{X}_1^{\pi/2}$.
    \textbf{d}, Full error generator for device B, $\text{X}_1^{\pi/2}$.
    \textbf{e}, Full error generator for device A, $\text{X}_2^{\pi/2}$.
    \textbf{f}, Full error generator for device B, $\text{X}_2^{\pi/2}$.}
    \label{extFig:GSTerrorgenerators}
\end{figure*}

\fi

\end{document}